\documentclass[12pt,superscriptaddress,nofootinbib]{revtex4-2}

\usepackage[english]{babel}

\usepackage[a4paper,top=2cm,bottom=2cm,left=3cm,right=3cm,marginparwidth=1.75cm]{geometry}
\usepackage{setspace}
\usepackage{amsmath}
\usepackage{graphicx}
\usepackage[colorlinks=true, allcolors=blue]{hyperref}
\usepackage{braket}

\begin{document}
	\title{Revivals and quantum carpets for the relativistic Schr\"odinger equation}
    \author{B. Zumer\footnote[1]{Current affiliation: Physikalisches Institut, Albert-Ludwigs-Universität Freiburg, Germany}}
    \affiliation{Laboratoire de Physique Théorique et Modélisation,
	CNRS Unité 8089, CY Cergy Paris Université,
	95302 Cergy-Pontoise cedex, France}

    \author{F. Daem}
    \affiliation{Laboratoire de Physique Théorique et Modélisation,
	CNRS Unité 8089, CY Cergy Paris Université,
	95302 Cergy-Pontoise cedex, France}
    
    \author{A. Matzkin}
    \affiliation{Laboratoire de Physique Théorique et Modélisation,
	CNRS Unité 8089, CY Cergy Paris Université,
	95302 Cergy-Pontoise cedex, France}
	
	\begin{abstract}
		
		We investigate wave packet dynamics for a relativistic particle in a box evolving according to the relativistic Schr\"odinger (also known as the Salpeter) equation. We derive the solutions for an infinite well -- which contrary to the standard relativistic wave equations (such as the Klein-Gordon or Dirac equations) -- are well defined, and use these solutions to construct wave packets. We obtain expressions for the wave packet revival times and explore the corresponding quantum carpets (the space-time probability density plots) for different dynamical regimes. We further analyze level spacing statistics as the dynamics goes from the non-relativistic regime to the ultra-relativistic limit.
	\end{abstract}
	
	\maketitle

	\section{Introduction}
	
	The dynamics of an initially localized wave packet of a quantum particle displays, even in simple systems, very interesting features. In particular, wave packet revivals \cite{robinett} is a striking effect by which, after initially spreading throughout all available configuration space, a wave packet relocalizes (entirely or partially). In non-relativistic systems, wave packet revivals were much investigated theoretically \cite{averbukh,Aronstein,carpet-ori1,carpet-ori2,Buchleitner}, and observed experimentally \cite{exp1,exp2,exp3} in some systems.   
	
	Extensions to relativistic systems are scarce. Early works explored slightly relativistic regimes \cite{Schleich,early}. A mathematically interesting (but non-physical) case of a Dirac particle constrained on a circle was also studied \cite{Strange,dirac-revivals}. The reason is that high (so called supercritical) potentials are non-binding in the relativistic domain. For instance the non-relativistic system that has been used as the main support to investigate theoretically the properties of revivals has been the particle in a box (an infinite well) \cite{robinett,Bluhm,carpet-ori1,carpet-ori2,Aronstein,Styer,heller}, given that the eigenfunctions have a simple form and are easily tractable analytically. However the relativistic infinite well is an intricate problem \cite{rqm_well,alkhateeb-AJP,rqm_bc1,rqm_bc2}, due to the existence of Klein tunneling (whereby instead of being reflected on the walls of the box, a particle can leak outside the well as an antiparticle) and is therefore not well suited to the investigations of relativistic revivals.
	
	In this paper, we investigate revivals in an infinite well for a particle obeying the relativistic Schrödinger equation (RSE), also known in the literature as the Salpeter equation or the square-root Klein-Gordon equation, given by
	\begin{equation}
		i \hbar \partial_t \psi(x,t)= \sqrt{c^4m^2+c^2 \hat{p}^2} \, \psi(x,t)+V(\hat{x}) \, \psi(x,t) \label{eq:1}
	\end{equation}
	Contrary to the standard relativistic wave equations such as the Klein-Gordon or the Dirac equations, the Salpeter equation does not describe antiparticles and hence does not give rise to Klein tunneling. The RSE has been employed as a phenomenological tool to investigate low-energy relativistic phenomena for spinless or spin-averaged particles, in particular the bound states of hadrons \cite{meson-app-2003,meson-app-2006}, and has has recently been the object of renewed interest\cite{multiple-deltas-2017,eckstein,pavsic,annalen,us-dynamics,daem-tunneling}. Coupled RSEs also appear in the Foldy-Wouthuysen representation of the usual Klein-Gordon or Dirac equations \cite{daem-superradiance}.
	
	The main difficulty in solving Eq.\eqref{eq:1} lies in the presence of the square-root in the Hamiltonian, which becomes the square-root of a differential operator in configuration space. We will show in Sec. \ref{sec:energies} how to obtain the time-independent solutions of the RSE for an infinite well by going to momentum space. We will then define the wave packet construction and the revival times; we will in particular prove a conjecture concerning the revival period made in Ref. \cite{Schleich} on the basis of results obtained in the slightly relativistic regime. In Sec. \ref{sec:carpet} we will compute the wave packet evolution for the three typical regimes: non-relativistic limit, ultra-relativistic limit, and the intermediate regime. The dynamics will be displayed in terms of quantum carpets, a space-time plot showing the probability density. We will see that the ridges and canals that appear in non-relativistic quantum carpets take here characteristic forms. These results will be discussed and compared to previous results in Sec. \ref{sec:discussions}.
	
\section{Analytical solutions and revival times}

\subsection{Eigenfunctions of the 1D relativistic Schr\"{o}dinger equation in
	a well}

\label{sec:energies}

We first need to find the energy spectra and the corresponding eigenfunctions
of the Salpeter problem in a well. Due to the presence of the square root of a
differential operator in Eq. \ref{eq:1}, this cannot be carried out directly
in position space.\ It is usual \cite{Kowalski} instead to work in momentum space
where the Salpeter equation is easier to handle. Eq. \eqref{eq:1} becomes%

\begin{equation}
	i\hbar\partial_{t}\psi(p,t)=\sqrt{m^{2}c^{4}+p^{2}c^{2}}\psi(p,t)+\frac
	{1}{\sqrt{2\pi\hbar}}\int \mathrm{d}p^{\prime}\tilde{V}(p-p^{\prime})\psi(t,p^{\prime
	}) \,, \label{eq:2}%
\end{equation}
where%
\begin{equation}
	\tilde{V}(p-p^{\prime})=\frac{1}{\sqrt{2\pi\hbar}}\int dxV(x)e^{-ix(p^{\prime
		}-p)/\hbar} \label{vft}%
\end{equation}
is the Fourier transform of the potential. Separating the variables leads to
solutions of the form $\xi_{n}(t,p)=\exp\left(  -i E_{n}t/\hbar\right)
\phi_{n}(p)$ where $\phi_{n}(p)$ obeys%
\begin{equation}
	E_{n}\phi_{n}(p)=E(p)\phi_{n}(p)+\frac{1}{\sqrt{2\pi\hbar}}\int
	\mathrm{d}p^{\prime}\tilde{V}(p-p^{\prime})\phi_{n}(p^{\prime}). \label{teie}%
\end{equation}
In general, finding solutions in closed form of this integral equation is
impossible, and Eq. \eqref{teie} must be solved numerically (e.g.
\cite{daem-tunneling}).

In the present case, we are considering a one-dimensional box potential
between $0$ and $L$ that can be taken as the $V_{0}\rightarrow\infty$ limit of
the finite well potential defined by $V(x)=V_{0}\left[\theta
(-x)+  \theta(x-L)\right]$, where $\theta$ is the unit step function. $\tilde{V}(p)$ can
be straightforwardly obtained from Eq. \eqref{vft}, and Eq. \eqref{teie} can
be written in the form%
\begin{equation}
	\phi_{n}(p)=\dfrac{1}{\sqrt{2\pi}}\frac{\int \mathrm{d}p^{\prime}\mathcal{V}(p-p^{\prime})\phi_{n}(p^{\prime
		})}{\frac{E_{n}-E(p)}{V_{0}}-1}\label{simp}%
\end{equation}
with $\mathcal{V}(p)=i\left(  1-e^{-iLp}\right)  /\left(  \sqrt{2\pi}p\right)
$ and $\hbar=1$. In the limit $V_{0}\rightarrow\infty$, Eq. \eqref{simp} implies that the
solutions $\phi_{n}(p)$ are independent of the specific form of the kinetic
term $E(p)$, and we can therefore expect the eigenfunctions and the
quantization condition for the Salpeter equation to be identical to those of
the standard non-relativistic particle in a box problem. Indeed, by solving
the integral equation \eqref{simp}, one finds (see Appendix \ref{sec:annexA}) that up to a
normalization factor

\begin{equation}
	\phi_{n}(p)=\frac{\left(  -(-1)^{n}+e^{-iLp}\right)  }{p^{2}-k_{n}^{2}%
	}\label{solfi}%
\end{equation}
with $k_n=n\pi/L$, $n$ being an integer. The Fourier transform of Eq.
\eqref{solfi} gives the familiar $\phi_{n}(x)=\sin(k_{n}x)$ inside the box, and the
corresponding eigenvalues are%

\begin{equation}
	E_{n}=\sqrt{m^{2}c^{4}+k_{n}^{2} \, c^{2}}=mc^{2}\sqrt{1+n^{2} \, \lambda_{C}%
		^{2}/(2L)^{2}}\,,\quad n\in\mathbf{N^{\ast},}\label{eq:E}%
\end{equation}
where $\lambda_{C}=h/mc$ is the Compton wavelength.

As expected, the energy eigenvalues tend to the non-relativistic ones
obtained by solving the Schr\"{o}dinger equation (up to the constant
$mc^{2}$) for low values of $k_n$ as it shows a quadratic dependence of
the energy with $n$.

In the other extreme (ultrarelativistic
regime), for very large $k_n$, the energy scales linearly with $n$, as
for an harmonic oscillator. Note that the size of the box has a direct impact
on the dynamics contrary to the Schr\"{o}dinger case, where everything can be
scaled by $L$, here we have a natural unit of length that is the Compton
wavelength. It is to be directly compared to the length of the periodic orbit
that is twice the size of the box. For small (relative to the Compton
wavelength) boxes, we are thus immediately in the relativistic regime, even
for $n=1$).

Note that quantization did not appear by imposing Dirichlet boundary
conditions in configuration space, but by a compatibility condition in
momentum space.\ This compatibility condition, requiring that Eq. (\ref{teie})
remains finite as $V_{0}\rightarrow\infty$ is independent of the form of the
kinetic energy, so that the ansatz made in  Appendix \ref{sec:annexA} is
rather straightforward.

	\subsection{Classical and quantum relativistic revival time}
	\label{sec:revival}
	Now that we have established the energy spectrum for the relativistic Schrödinger particle in a box, one may look at the revival time for an initial wave packet placed in the well. To do so, it is important to consider wave packets that are localized with their energy spectrum not too spread around a central value $n_0$ \cite{robinett} that represents the dominant contribution. This assumption is well verified in the case of Gaussian wave packets in the form:
	
	\begin{equation}
		\Psi(x, 0) = A \exp\left( -\frac{(x - x_0)^2}{4\sigma^2} + \frac{i p_0 x}{\hbar} \right)
	\end{equation}
	whose time evolution is simply given by
	
    \begin{equation}
        \Psi(x,t)= \sum_n a_n e^{-i E_n t/\hbar}\,,
    \end{equation}
    where the initial wave packet is decomposed in the energy eigenbasis as
    
    \begin{equation}
        \ket{\psi (t=0)} = \sum_n a_n \ket{E_n}\,.
        \label{eq:decomposition}
    \end{equation}
    
	By doing so, and rewriting the energy eigenvalues of the system as
	
	\begin{equation}
		\begin{split}
			E(n) \approx \, & \, \, \,\,E(n_0) +E^\prime(n_0) (n-n_0) \\
			& + E^{\prime\prime}(n_0) \dfrac{(n-n_0)^2}{2} +  E^{\prime\prime\prime}(n_0) \dfrac{(n-n_0)^3}{6} +  ...
		\end{split}
		\label{eq:E_decompose}
	\end{equation}
	
	We can rewrite the time dependence of the considered wave packet as
	
	\begin{equation}
		\begin{split}
			e^{-iE_nt/\hbar}= & \,\,\, \text{exp}\left( -i \omega_0t-2\pi i  (n-n_0) t/T_{cl}  \right. \\
			& \left. - 2\pi i(n-n_0)^2 t/T_{rev}- 2\pi i(n-n_0)^3 t/T_{super}+...                    \right)\,,
		\end{split}
		\label{eq:time}
	\end{equation}
	showing clearly the appearance of revival times of different orders.
	This relation between the energy and the revival times have already been well studied \cite{robinett}.
	
	The first time appearing, $T_{cl}$, is the classical revival. This one corresponds to the time it take for a classical particle to perform one full period of the trajectory, where the classical particle obeys Hamilton's equation of motion for the classical non covariant Hamiltonian \cite{barut2010} $\sqrt{c^4m^2+c^2 p^2}+V(x)$.
	This time can be expressed as
	
	\begin{equation}
		T_{cl}=\dfrac{2\pi \hbar}{|E_{n_0}^\prime|} \,.
	\end{equation}
	
	In a non relativistic billiard, this is known to be
	
	\begin{equation}
		T_{cl}=\dfrac{2L}{v_0}\,,
	\end{equation}
	which is simply the size of the periodic orbit divided by the (here constant) speed at which the orbit is spanned.
	
	And indeed, when calculating the derivative of the energy with respect to $n$, and using the fact that $E=\gamma m c^2$ and $p_n = \hbar n\pi/L = \gamma m v_n$, the relativistic classical time is given by
	
	\begin{equation}
		T_{cl}^{R} = \frac{2\pi\hbar}{|E_{n_0}^\prime|} = \frac{2L}{v_{n_0}}\,,
	\end{equation}
	which is the same expression as for the non relativistic case except that $v_n$ is now the relativistic velocity. Hence, the revival time is constrained indirectly by the bound $c \geq v_n$. 
	This can be seen in the high energy limit  where the above expression becomes
	
	\begin{equation}
		T_{cl}^{R} \xrightarrow[n_0 \to \infty]{} \frac{2L}{c}\,.
	\end{equation}

	A second characteristic time for the system is the quantum revival time. This corresponds to the spread of the wave packet for longer times. Here, the spread of the wave packet has nothing to do with the classical dynamics anymore but relies on the quantum nature of the system as we observe the wave packet interfere with itself. 
	
	This revival time  can be expressed as:

	\begin{equation}
		T_{rev} = \frac{2\pi\hbar}{|E_{n_0}^{\prime \prime}|/2} \,,
	\end{equation}	
		
	For the non non relativistic quantum well problem, the revival time is given by
	\begin{equation}
		T_{rev}^{NR} = \frac{4mL^2}{\pi\hbar} = (2n_0) \, T_{cl}^{NR}\,.
	\end{equation}

	For the Salpeter problem, the relativistic revival time is given exactly in the simple form
	\begin{equation}
		T_{rev}^{R} = (2n_0)\, \dfrac{2L}{v_{n_0}} \gamma_0^2 = T_{rev}^{NR} \gamma_0\,.
	\end{equation}
	This proves a conjecture made in \cite{Schleich} on the basis of the Schr\"odinger equation with a first order relativistic correction.
	
	It is interesting to notice that this revival time tends to infinity when the velocity increases. This means that in the ultra-relativistic regime, we expect no quantum revivals and thus, very limited interference effects.
	This can be seen quite clearly when looking at equation \eqref{eq:E} where we see that the energy spectra becomes almost linear when $n$ is large enough. This linear dispersion relation signals the presence of non dispersive wave packets \cite{Buchleitner}.

	Finally, it can be of interest to compute the next typical time in the expansion \eqref{eq:time}. This so called ``super-revival time'' does not appear in the non-relativistic particle in a box as it involves the third derivative of the energy with respect to $n$:
	\begin{equation}
		T_{super} = \frac{2\pi\hbar}{|E_{n_0}^{\prime \prime \prime}|/6} \,,
	\end{equation}
		Because it involves the third derivative of $E(n)$, it diverges to infinity both in the ultra relativistic and non relativistic regimes. Thus, it will only be relevant in the intermediate regime between ultra-relativistic and non-relativistic dynamics. 
	
\begin{figure}[ht!]
		\centering
		\includegraphics[width=\linewidth]{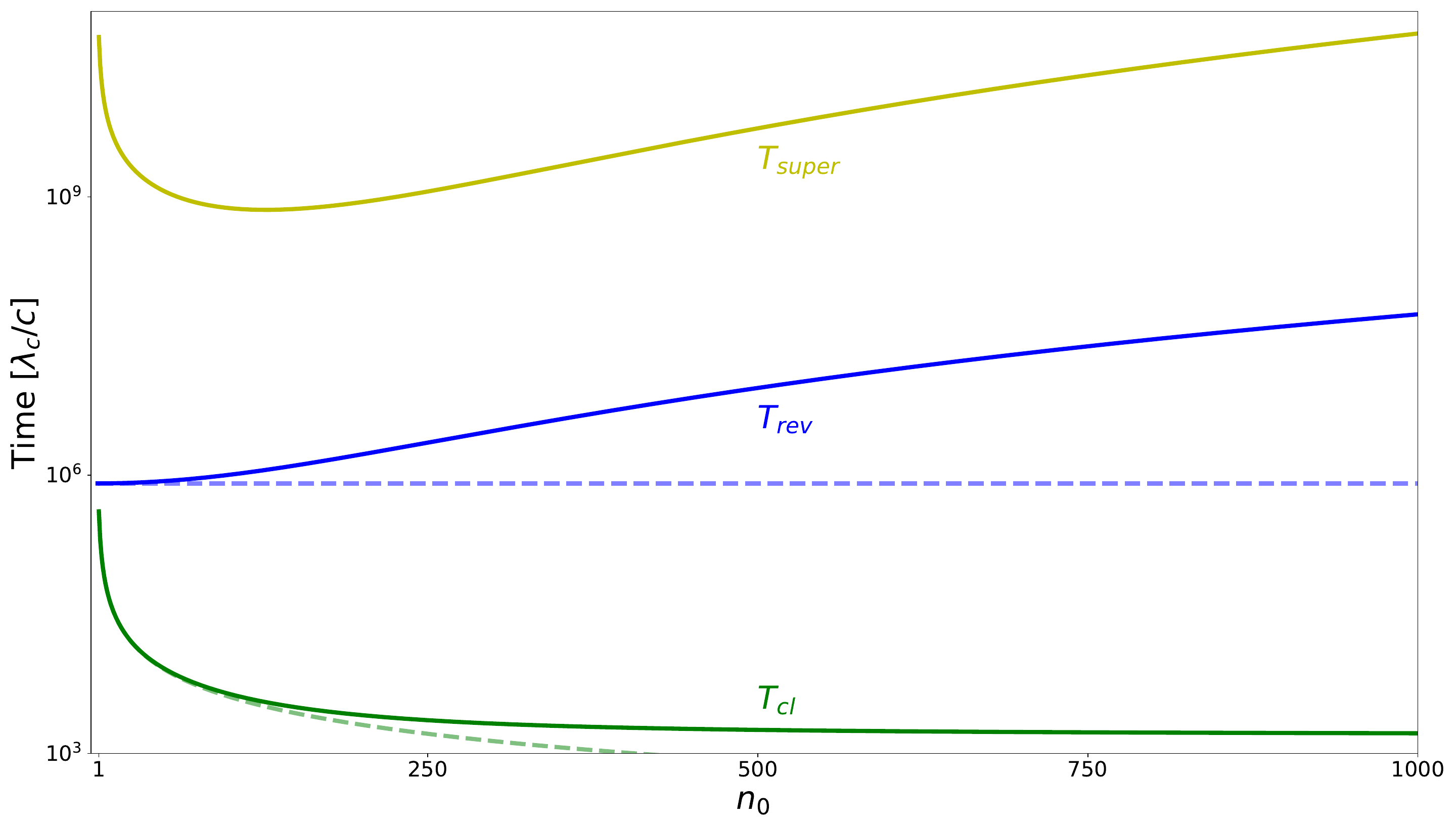}
		\caption{ Variation of the different revival times for the Salpeter particle in a box for $L=800 \, \lambda_C$. We are able to identify two regimes, the low energy regime ($n_0$ between 1 and about 50) where the revivals times corresponds to the non-relativistic particle in a box (as shown in dashed lines) and the high energy regime (around $n_0$ greater than 500) where the velocity tends to $c$.}
		\label{fig:times}
	\end{figure}

	For the Salpeter particle in a box, we can write this time as:
	\begin{equation}
		T^{R}_{super} =n_0 \:  T^{R}_{rev}\, \, \,  \dfrac{c^2}{v_0^2}\,,
		\label{eq:T_super}
	\end{equation}
	From the above expression, it is clear that $T^{R}_{rev} \ll T^{R}_{super}$. As shown in Appendix \ref{sec:annex} this trend continues and the next revival times keep on increasing.
	The different scales of these different revival times and their variation as a function of the energy is illustrated in a typical case in Fig. \ref{fig:times}.
	In this figure, we observe in particular the fact that for low $n_0$ the classical revival time decreases as $1/n_0$ as for the  Schr\"odinger equation. For large enough values of $n_0$, the effect of the light cone sets in,  giving a limit on the velocity of the particle and thus explaining the plateau of $T_{cl}$ for large $n_0$.
	On the other hand, we observe that $T_{rev}$ is not constant anymore given that the second derivative of $E(n)$ increases with $n$ (recall that $E^{\prime \prime}(n)$ is constant for the Schr\"odinger equation).
	Finally, higher order derivatives of $E(n)$ do not vanish either meaning in particular that $T_{super}$ starts by decreasing from infinity at low $n_0$ and then increases according to equation (\ref{eq:T_super}).

	\section{Quantum carpets for the Salpeter one dimensional billiard}
	\label{sec:carpet}
	\subsection{Non-relativistic regime: Schrödinger like revivals}
	\label{sec:non-relat_carpet}
\begin{figure}[ht!]
		\centering
		\includegraphics[width=\linewidth]{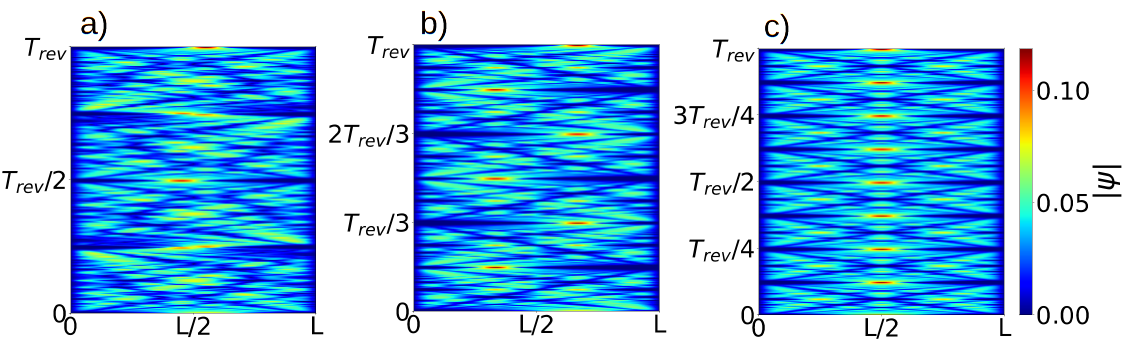}
		\caption{Quantum carpet for the Salpeter equation in a well. All initial Gaussian wave packet start with $\Delta x= L/20$ and no initial velocity. Yet, depending on the initial position, we observe new revivals. Figure (a) starts with a wave packet having no clear symmetry and thus we simply observe regular revivals. On the other hand, figure (b) and (c) starts with Gaussian centered at $2L/3$ and $L/2$ respectively thus exhibiting new revivals at fractions of the usual revival time. }
		\label{fig:carpets_non_relat}
	\end{figure}

	In the low velocity regime, the Salpeter equation behaves like the Schrödinger equation. Hence, we expect to find usual revivals as described by Robinett \cite{robinett}.
	And indeed, we are able to find revivals for the Salpeter equation in the low energy regime. This is best shown in Fig. \ref{fig:carpets_non_relat} in the form of quantum carpets --  a space-time plot showing the probability density, revealing a tapestry-like pattern characterized by ridges and canals arising from wave packet interference and revivals.
	
	Fig. \ref{fig:carpets_non_relat} displays a Gaussian wave packet at three different initial positions. These normalised Gaussian wave packets are then evolved in time using the split-operator method (see section \ref{sec:Num}).  All of these wave packets start with zero initial velocity and have a width in the momentum space of about $3.10^{-2} \, \hbar/L$, and hence a mean energy of $(10^{-4}+1) mc^2$ well in the non-relativistic regime. The corresponding quantum revival time here is $10^3 L/c$. Note that in the three panels shown in Fig. \ref{fig:carpets_non_relat}, only the initial position in the well changes. We observe that for a wave packet centered in the well, we have much more revivals, appearing for quarters of the revival time of the system at this energy. The same holds  for a wave packets starting at two thirds of the well, we observe more revivals than for a wave packet starting with no clear symmetry inside the well. Note that if the Gaussian width is increased, the positions of the canals and revivals are not modified, but they will be wider and thus more blurred.
	
	All of these quantum carpets show the usual structure of ridges and canals enhancing the wave-like propagation of the wavefunction density with constructive interference leading to high probability of finding the particle at the fringes.
	
	The revival time has been calculated from the energy spectra of the problem and obtained in Sec. \ref{sec:revival}.  It coincides perfectly with the numerical calculations we performed.
	
	\subsection{Relativistic regime: only classical revivals}
	
	Let us now  explore the ultra-relativistic regime. To do so we used two approaches; either we considered very narrow Gaussian wave packets with $\Delta k$ up to $15 \hbar/L$ or we considered the same wave packets as in the previous section \ref{sec:non-relat_carpet} but with a high initial velocity.
	
\begin{figure}[ht!]
		\centering
		\includegraphics[width=\linewidth]{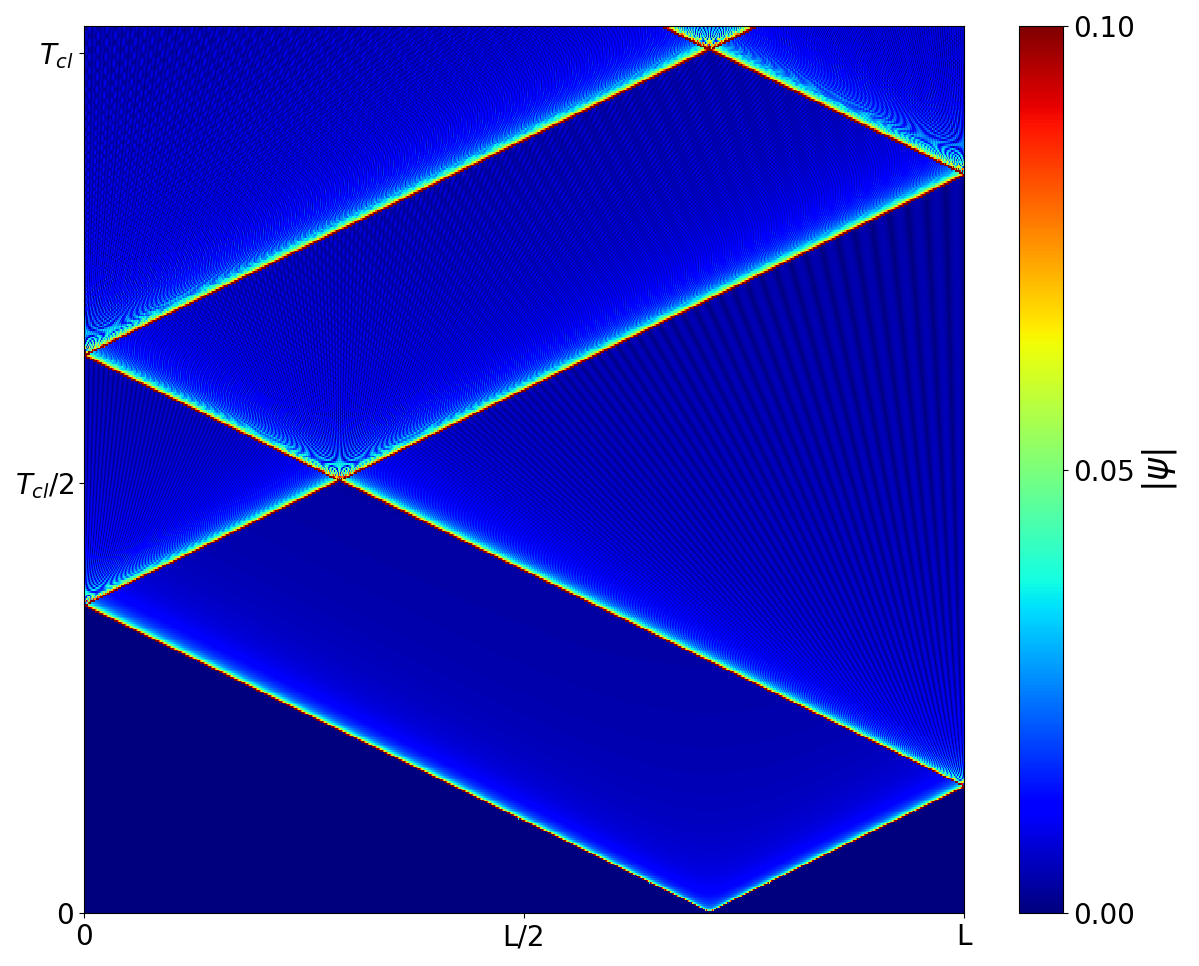}
		\caption{Propagation of a wave packet in the relativistic regime. The propagation is affected by the light cone as the wave packet cannot spread beyond it. As a consequence the interference effects are minimal here, the wave packet  bounces in a classical-like manner on the walls of the well. The typical revival time is simply given by the classical time $T_{cl}=2L/v$ which is here close to $2L/c$.}
		\label{fig:relat_psi}
	\end{figure}

	In both cases, we observe that the wave function is constrained by the light cone as shown on Fig. \ref{fig:relat_psi} where we display the propagation of a very narrow Gaussian wave packet ($\sigma_x=10^{-5} L$) and no initial velocity ($p_0=0$) .
	
	In the high energy regime, there is almost no dispersion effect meaning that the quantum revival time becomes extremely large. Hence, we simply see classical revivals appearing for times $T_{cl}$ that are close to $2L/c$. The particles propagate as expected for a classical particle; simply bouncing in the box. These behaviours are the one expected from our calculations carried in Sec. \ref{sec:revival}.

	\subsection{Intermediate regime: interplay between classical and quantum revivals}
	
\begin{figure}[ht!]
		\centering
		\includegraphics[width=\linewidth]{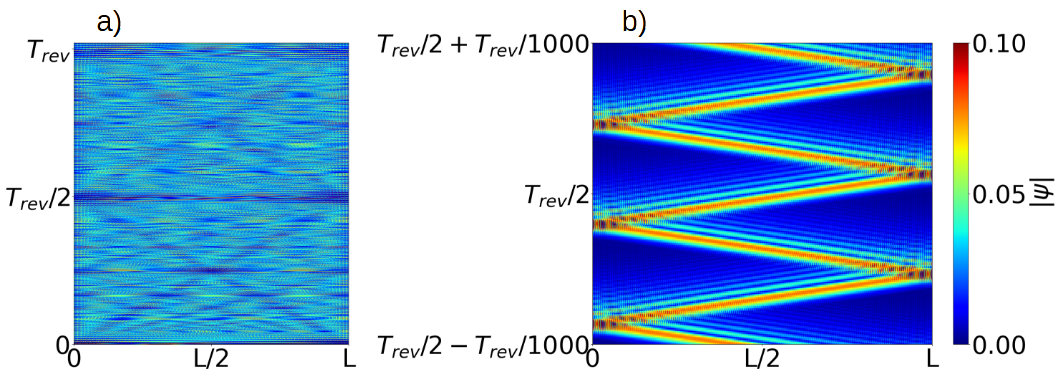}
		\caption{Quantum carpet in the intermediate regime with a Gaussian wave packet starting in the middle of the box with an initial velocity of $p_0=1.2 \hbar/L$ and width of $L/25$. We observe in (a) the appearance of interference patterns forming the usual ridges and canals but blurred out. Figure (b) shows a zoom of the quantum carpet around $t=T_{rev}/2$. On the zoom, we see that the interference pattern is actually the effect of many rebounds corresponding to the particle bouncing inside the well.}
		\label{fig:carpet-mix}
	\end{figure}

	We have discussed the two extreme regimes, the Schr\"odinger limit and the ultra relativistic regime. Now, one can wonder what happens at the interplay between the two. In order to do so, we consider again a Gaussian wave packet wide in position space ($\sigma_x=4\, 10^{-2} L$) with initial velocity such that the typical energy is located in the intermediate regime ($p_0=1.2 \hbar/L$). By doing so, we have a quantum revival time that is much larger than the classical revival time, which is already very close to  $2L/c$. Typically, the ratio of $T_{rev}/T_{cl}$ is of the order of $1500$. This means that the time it takes for interference to occur is extremely slow in comparison to the classical time. Hence, in Fig. \ref{fig:carpet-mix}, the quantum carpet appears on top of the very fast oscillations of the wave packet in the well.

	\section{Discussions}
	\label{sec:discussions}
	
	\subsection{Revival times in the relativistic regime}
	
	Revival times have seldom been discussed for relativistic models. An exception is in a massive Dirac particle constrained to move on a ring \cite{Strange,Chamizo}. On the other hand, in a slightly relativistic situation (that can be seen as an approximation of the Salpeter equation), quantum revivals have been observed \cite{Schleich}. Nevertheless, it was difficult to observe them as this required fine tuning.
	
	Here we have shown the existence of revivals for any regime in this relativistic model. We derived revival times of any order (see Appendix \ref{sec:annex}) in the Salpeter model valid for any value of the initial velocity. As the series given, by equation (\ref{eq:E_decompose}) is infinite, we have no complete revival in contrast to the non-relativistic case. This means that the pattern of canals and ridges will start to blur after a sufficiently long time due to the effect of the higher order terms in equation (\ref{eq:time}). Finally, we linked these revival times to classical periods highlighting the appearance of purely relativistic terms.

	\subsection{Numerical approaches}
	\label{sec:Num}
	The relativistic Schr\"odinger equation in an infinite well is special in that the eigenfunctions are known
	analytically. However this will not be the case for other systems that will require numerical solutions of
	integral equations in momentum space \cite{daem-tunneling}. Having in mind applications to billiards 
	of arbitrary shape in the context of relativistic quantum chaos, the relativistic particle in a box can be used
	as a benchmark to test purely numerical approaches. Here we have employed two approaches.
	
	First, following the method used in \cite{daem-tunneling}, a diagonalization of the Hamiltonian can be performed in momentum space by discretizing Eq. \eqref{teie} in the form of a matrix equation and solving for all eigenvalues and discretized eigenfunctions at once. Evolution of any initial wave packets are then readily obtained at any time after decomposing them on the basis of eigenfunctions.
	
	Second, to obtain the density plots presented in this article, we used a more direct split operator method that applies a discretized time-evolution operator to the initial wavefunction (hence without computing the eigenvalues).
	In order to extract the energy levels of the system we compute the Fourier transform of the auto correlation function. Indeed, the auto-correlation function is linked to the energy levels through
	
	\begin{equation}
		A(t)= \int \psi^*(t=0,x)\psi(t,x) \mathrm{d}x = \braket{\psi(t=0)|\psi(t)} = \sum_n |a_n|^2 e^{-i E_n t/\hbar}\,,
	\end{equation}
    where the coefficient $a_n$ comes from the decomposition \eqref{eq:decomposition}
	
	Hence, by letting a Gaussian wave packet evolve for a long time, we resolve the energy levels of the system. To do so we consider narrow Gaussian wave packets in order to populate a wide range of energy levels. In order to access even higher energy levels, we can also use narrow Gaussian wave packet with high initial velocity, hence shifting the population of energy levels by a fixed energy corresponding to $p_0$.
	%While using this method, one has to be careful as systematic extinction could append \cite{robinett}.
	
	The split operator method is more efficient and scalable to compute the evolution of wavefunctions and their associated probability density. However, a direct diagonalization of the hamiltonian, despite being expensive and less applicable to multidimensional systems, can be useful if a more precise determination of the coefficients $a_n$ and energies $E_n$ is required.
	
	\subsection{Symmetries and extinction of coefficients}
	
\begin{figure}[h]
		\centering
		\includegraphics[width=0.4\linewidth]{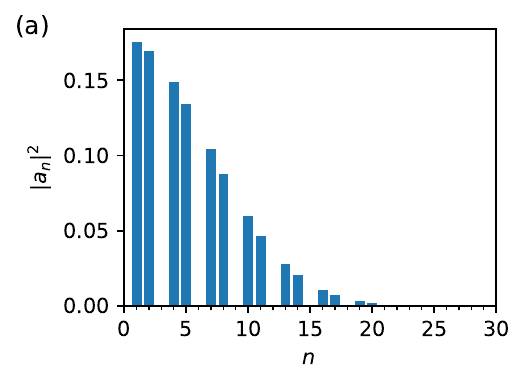}
		\includegraphics[width=0.4\linewidth]{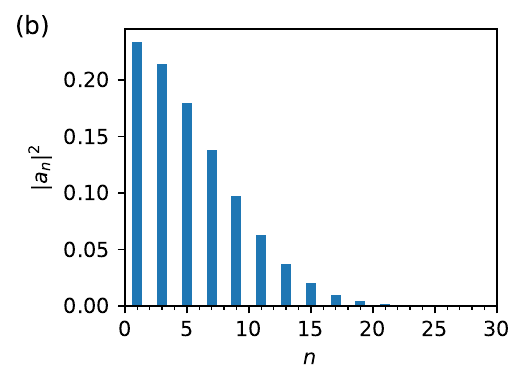}
		\caption{Population distribution $|a_n|^2$ for an initial Gaussian wave packet with $p_0=0$. (a) $x_0=2L/3$, (b) $x_0=L/2$. In both cases, the coefficient distribution follows a Gaussian curve centered at $n=0$. Due to the symmetry of the initial wave packet in the well, a third (a) or a half (b) of the coefficients vanish. The coefficients are computed using the diagonalization method.}
		\label{fig:Extinction}
	\end{figure}

	Due to symmetries in the problem, fractional positions $x_0$ of the initial wave packet lead to extinction of some of the coefficients $a_n$ in the decomposition \eqref{eq:decomposition}. For a Gaussian packet, the expression of these coefficients is known with the assumption that the tails have negligible contributions outside of the box \cite{robinett}. Since the eigenfunctions of the infinite well problem are identical in the relativistic or non relativistic Schrödinger equation, the coefficients are the same for a given Gaussian and give the same extinctions, as shown in Fig. \ref{fig:Extinction}.
	
	As a way of validating our numerical methods, the coefficients in Fig. \ref{fig:Extinction} are obtained using the diagonalization method. Although this numerical approach is not necessary in our specific case, as we know the analytical solutions, it will be important when considering two dimensional billiards of any shape. Indeed, we can generalise this approach to any potential -- smooth or not -- of our choice.
	
	\subsection{Energy spectra and level spacing}
	
	Since we can determine the energy spectra (both analytically and numerically), it is interesting to look at the level spacing. 
	Indeed, the Salpeter equation displays a peculiar  spacing between nearest neighbors that depends on the regime. The spacing scales as $n$ in the low energy regime, starting with energy differences close to zero as expected for the Schrödinger infinite well. But on the other hand, in the high energy regime, the dispersion relation tends to be linear, hence the energy levels are always separated by almost the same value $\hbar \pi c/L=mc^2 \lambda_C/(2L)$. A transition between both behaviors can be seen in the intermediate regime.
	
\section{Conclusion}

We have investigated the dynamics of a relativistic spinless particle confined in a one-dimensional infinite potential well using the Salpeter equation. We have demonstrated that the Salpeter equation allows for well-defined wave packet revivals in an infinite well, unlike other relativistic models that suffer from complications such as Klein tunneling. Despite the non-local nature of the relativistic Hamiltonian, we found that the eigenfunctions remain identical to those of the non-relativistic case, while the energy spectrum is significantly altered, especially in the ultra-relativistic regime.

By examining the time evolution of initially localized wave packets, we observed distinct behaviors across different energy regimes. In the non-relativistic limit, the system exhibits familiar features such as quantum revivals and fractional revivals, reflecting quantum coherence. In contrast, the ultra-relativistic regime leads to a linear energy spectrum, which suppresses these revival structures and results in motion that resembles classical particle dynamics.

These results provide a clear illustration of how relativistic corrections
influence quantum systems even in simple geometries. They could be useful to
model the propagation of confined relativistic particles below the
supercritical regime, or when investigating optical analogues of relativistic
quantum particles. Overall, our results highlight the Salpeter equation as a
consistent and insightful framework for exploring relativistic quantum
dynamics in bounded systems, bridging the gap between quantum and classical behavior.

	\appendix
	
		\section{Integral equation for the eigenfunctions}
	\label{sec:annexA}
	
	In the $V_{0}\rightarrow\infty$ limit, Eq.\ \eqref{simp} becomes
	\begin{equation}
		\phi(p)=\frac{1}{2\pi i}\int dp^{\prime}\frac{1-e^{-iL\left(
				p-p^{\prime}\right)  }}{p-p^{\prime}}\: \phi(p^{\prime}). \label{inteqlim}%
	\end{equation}
	Let us make an educated guess by looking for $\phi$ in the form $\phi
	(p)=\left(  a\exp\left(  -ipL\right)  +b\right)  /\left(  p^2-k^2\right)$.
	Since we have simple poles at $\pm k$,  Eq. \eqref{inteqlim}
	gives
	\begin{equation}
		\frac{1}{2k}\left[  \left(  a\exp\left(  -ikL\right)  +b\right)
		\frac{\left(  1-e^{-iL(p-k)}\right)  }{p-k}-\left(  a\exp\left(  +ikL\right)
		+b\right)  \frac{\left(  1-e^{+iLk}\right)  }{p+k}\right]  .
	\end{equation}
	Note now that this expression has removable poles, while $\phi(p)$ has poles
	in the neighborhood of $p=\pm k$. Therefore the poles of $\phi(p)$ at $p=\pm k$ must vanish which is only possible provided $b=-a\exp(ikL)\ $with $k\equiv k_{n}=n\pi/L$. By replacing these values in the ansatz for $\phi(p)$, one obtains the solution
	given by Eq. \eqref{solfi}.
	
	\section{Computation of the energy derivatives}
	\label{sec:annex}
	
	The formula for derivatives of order $N$ of $E_n=\sqrt{m^2c^4+(\hbar\pi c /L)^2n^2}$ are obtained after a factorisation of $mc^2$ and use of a tabulated general expression for functions of the form $f(x)=(1+ax^2)^p$ \cite{gradstejn_table_2009}. Using the relations $p_n=\hbar\pi n/L$ and $E_n=\gamma mc^2$ the equation is then readily expressed in terms of $p_n$, $E_n$, and $\gamma$, we  get
	
	\begin{equation}
		\frac{\mathrm{d}^N E_n}{\mathrm{d}n^N}
		= \frac{\left(2 \left(\frac{\hbar\pi}{L}\right) p_n\right)^N}
		{c^{2N-2} (\gamma m)^{2N-1}} \prod_{j=0}^{N-1} (1/2-j)
		\sum_{k=0}^{\lfloor N/2 \rfloor} \frac{\prod_{r=1}^{2k}(N-r)}{k!\prod_{q=1}^{k}(1/2-N+q)}
		\left(\frac{E_n}{2p_n c}\right)^{2k}
		\label{eq:T_general}
	\end{equation}
	
	Note that the first term in the sum over $k$ ($k=0$) is always equal to one.
	Thus, the expression for the first derivative is trivially
	
	\begin{equation}
		E_n^\prime = \frac{\hbar\pi}{L} \frac{p_n}{\gamma m}\,.
	\end{equation}
	
	For the second derivative, the sum has a second term and the result simplifies to
	
	\begin{equation}
		E_n^{\prime\prime} = \left( \frac{\hbar\pi}{L} \right)^2 \frac{1}{\gamma^3 m}\,.
	\end{equation}
	
	Computations of higher order derivatives become more involved as the number of terms in the sum increases and the physical meaning of the resulting expressions is not always straightforward. It is however relatively easy to compute the ratios of different energy derivatives and, by extension, ratios of revival times.

	\bibliographystyle{unsrt}
	\bibliography{bib}

@article{exp1,
  title = {Observation of fractional revivals in the evolution of a Rydberg atomic wave packet},
  author = {Yeazell, John A. and Stroud, C. R.},
  journal = {Phys. Rev. A},
  volume = {43},
  issue = {9},
  pages = {5153--5156},
  numpages = {0},
  year = {1991},
  month = {May},
  publisher = {American Physical Society},
  doi = {10.1103/PhysRevA.43.5153},
  url = {https://link.aps.org/doi/10.1103/PhysRevA.43.5153}
}

@article{exp2,
  title = {Observation of fractional revivals of a molecular wave packet},
  author = {Vrakking, Marc J. J. and Villeneuve, D. M. and Stolow, Albert},
  journal = {Phys. Rev. A},
  volume = {54},
  issue = {1},
  pages = {R37--R40},
  numpages = {0},
  year = {1996},
  month = {Jul},
  publisher = {American Physical Society},
  doi = {10.1103/PhysRevA.54.R37},
  url = {https://link.aps.org/doi/10.1103/PhysRevA.54.R37}
}

@article{exp3,
title = {Real-time observation of vibrational revival in the fastest molecular system},
journal = {Chemical Physics},
volume = {329},
number = {1},
pages = {193-202},
year = {2006},
note = {Electron Correlation and Multimode Dynamics in Molecules},
issn = {0301-0104},
doi = {https://doi.org/10.1016/j.chemphys.2006.06.038},
url = {https://www.sciencedirect.com/science/article/pii/S0301010406003545},
author = {A. Rudenko and Th. Ergler and B. Feuerstein and K. Zrost and C.D. Schröter and R. Moshammer and J. Ullrich}}

@article{averbukh,
title = {Fractional revivals: Universality in the long-term evolution of quantum wave packets beyond the correspondence principle dynamics},
journal = {Physics Letters A},
volume = {139},
number = {9},
pages = {449-453},
year = {1989},
issn = {0375-9601},
doi = {https://doi.org/10.1016/0375-9601(89)90943-2},
url = {https://www.sciencedirect.com/science/article/pii/0375960189909432},
author = {I.Sh. Averbukh and N.F. Perelman}
}

@article{Chamizo,
	title = {Exact quantum revivals for the Dirac equation},
	author = {Chamizo, Fernando and Santill\'an, Osvaldo P.},
	journal = {Phys. Rev. A},
	volume = {109},
	issue = {2},
	pages = {022231},
	numpages = {12},
	year = {2024},
	month = {Feb},
	publisher = {American Physical Society},
	doi = {10.1103/PhysRevA.109.022231},
	url = {https://link.aps.org/doi/10.1103/PhysRevA.109.022231}
}

@article{Strange,
	title = {Relativistic Quantum Revivals},
	author = {Strange, P.},
	journal = {Phys. Rev. Lett.},
	volume = {104},
	issue = {12},
	pages = {120403},
	numpages = {4},
	year = {2010},
	month = {Mar},
	publisher = {American Physical Society},
	doi = {10.1103/PhysRevLett.104.120403},
	url = {https://link.aps.org/doi/10.1103/PhysRevLett.104.120403}
}

@article{Kowalski,
	title = {Salpeter equation and probability current in the relativistic Hamiltonian quantum mechanics},
	author = {Kowalski, K. and Rembieli\ifmmode \acute{n}\else \'{n}\fi{}ski, J.},
	journal = {Phys. Rev. A},
	volume = {84},
	issue = {1},
	pages = {012108},
	numpages = {11},
	year = {2011},
	month = {Jul},
	publisher = {American Physical Society},
	doi = {10.1103/PhysRevA.84.012108},
	url = {https://link.aps.org/doi/10.1103/PhysRevA.84.012108}
}

@article{robinett,
	title = {Quantum wave packet revivals},
	journal = {Physics Reports},
	volume = {392},
	number = {1},
	pages = {1-119},
	year = {2004},
	issn = {0370-1573},
	doi = {https://doi.org/10.1016/j.physrep.2003.11.002},
	url = {https://www.sciencedirect.com/science/article/pii/S0370157303004381},
	author = {R.W. Robinett}
}

@article{Schleich,
	author = {Marzoli, I. and Kaplan, A.E. and Saif, F. and Schleich, W.P.},
	title = {Quantum carpets of a slightly relativistic particle},
	journal = {Fortschritte der Physik},
	volume = {56},
	number = {10},
	pages = {967-992},
	doi = {https://doi.org/10.1002/prop.200810535},
	url = {https://onlinelibrary.wiley.com/doi/abs/10.1002/prop.200810535},
	year = {2008}
}

@article{Bluhm,
	author = "Bluhm, Robert and Kostelecky, V. Alan and Porter, James A.",
	title = "{The Evolution and revival structure of localized quantum wave packets}",
	eprint = "quant-ph/9510029",
	archivePrefix = "arXiv",
	reportNumber = "COLBY-95-05, IUHET-308, IUHET 308, July 1995",
	doi = "10.1119/1.18304",
	journal = "Am. J. Phys.",
	volume = "64",
	pages = "944--953",
	year = "1996"
}

@article{Styer,
	author = {Styer, Daniel},
	year = {2000},
	month = {01},
	pages = {},
	title = {Quantum revivals versus classical periodicity in the infinite square well},
	volume = {69},
	journal = {American Journal of Physics},
	doi = {10.1119/1.1287355}
}

@article{heller,
	title = {Semiclassical investigation of the revival phenomena in a one-dimensional system},
	author = {Zhe-xian Wang and Eric J. Heller},
	journal = {J. Phys. A: Math. Theor.},
	volume = {42},
	issue = {},
	pages = {285304},
	numpages = {},
	year = {2009},
	month = {Jun}
}

@article{Aronstein,
	title = {Fractional wave-function revivals in the infinite square well},
	author = {Aronstein, David L. and Stroud, C. R.},
	journal = {Phys. Rev. A},
	volume = {55},
	issue = {6},
	pages = {4526--4537},
	numpages = {0},
	year = {1997},
	month = {Jun},
	publisher = {American Physical Society},
	doi = {10.1103/PhysRevA.55.4526},
	url = {https://link.aps.org/doi/10.1103/PhysRevA.55.4526}
}

@article{carpet-ori1,
	author = "M V Berry",
	title = "{Quantum fractals in boxes}",
	eprint = " ",
	archivePrefix = "arXiv",
	journal = " J. Phys. A: Math. Gen.",
	volume = "29",
	pages = "6617",
	year = "1996"
}

@article{carpet-ori2,
	author = "Frank Großmann, Jan-Michael Rost and Wolfgang P Schleich",
	title = "{ Spacetime structures in simple quantum systems }",
	eprint = " ",
	archivePrefix = "arXiv",
	journal = " J. Phys. A: Math. Gen.",
	volume = "30",
	pages = "L277",
	year = "1997"
}

@article{alkhateeb-AJP,
	author = "M. Alkhateeb and A. Matzkin",
	title = "{Relativistic spin-0 particle in a box: Bound states, wave packets,
	and the disappearance of the Klein paradox}",
	eprint = " ",
	archivePrefix = "arXiv",
	journal = " Am. J. Phys.",
	volume = "90",
	pages = "297",
	year = "2022"
}

@article{rqm_well,
	author = {{P. Alberto, S. Das, and E. C. Vagenas}},
	title = "{Relativistic spin-0 particle in a box: Bound states, wave packets,
	and the disappearance of the Klein paradox}",
	archivePrefix = "arXiv",
	journal = "Eur. J. Phys.",
	volume = "39",
	pages = "025401",
	year = "2018"
}

@article{rqm_bc1,
	author = "Ar Rohim and Kazuhiro Yamamoto",
	title = "{Effects of chiral MIT boundary conditions for a
	Dirac particle in a box	}",
	eprint = " ",
	archivePrefix = "arXiv",
	journal = "Prog. Theor. Exp. Phys.",
	volume = "",
	pages = "113B01 ",
	year = "2021"
}

@article{rqm_bc2,
	author = "Salvatore De Vincenzo",
	title = "{General pseudo self-adjoint boundary conditions for a 1D KFG particle in a
	box
	}",
	eprint = " ",
	archivePrefix = "arXiv",
	journal = "Phys. Open",
	volume = "15",
	pages = "100151",
	year = "2023"
}

@article{early,
	author = "S. Ghosh and I. Marzoli",
	title = "{Super Revivals and Sub-Planck
	Scale Structures of a Slightly
	Relativistic Particle in a  Box
	
	}",
	eprint = " ",
	archivePrefix = "arXiv",
	journal = "Int. J. Quant. Inf.",
	volume = "9",
	pages = "1519",
	year = "2011"
}

@article{dirac-revivals,
	author = "F. Chamizo and O. P. Santillan",
	title = {Exact quantum revivals for the Dirac equation},
	eprint = " ",
	archivePrefix = "arXiv",
	journal = "Phys. Rev. A",
	volume = "109",
	pages = "022231",
	year = "2024"
}

@article{daem-tunneling,
	doi = {10.1088/1402-4896/ad9550},
	url = {https://dx.doi.org/10.1088/1402-4896/ad9550},
	year = {2024},
	month = {dec},
	publisher = {IOP Publishing},
	volume = {100},
	number = {},
	pages = {015216},
	author = {Daem, F. and Matzkin, A.},
	title = {Tunneling dynamics of the relativistic {Schr\"odinger/Salpeter} equation},
	journal = {Phys. Scr.},
}

@article{daem-superradiance,
	title = {Effects of superradiance on relativistic {Foldy-Wouthuysen} densities},
	author = {Daem, F. and Matzkin, A.},
	journal = {Phys. Rev. A},
	volume = {111},
	issue = {6},
	pages = {L060202},
	numpages = {5},
	year = {2025},
	month = {Jun},
	publisher = {American Physical Society},
	doi = {10.1103/d38q-h3qw},
	url = {https://link.aps.org/doi/10.1103/d38q-h3qw}
}

@article{meson-app-2003,
  title = {Reduction of the {QCD} string to a time component vector potential},
  author = {Allen, Theodore J. and Olsson, M. G.},
  journal = {Phys. Rev. D},
  volume = {68},
  issue = {5},
  pages = {054022},
  numpages = {10},
  year = {2003},
  month = {Sep},
  publisher = {American Physical Society},
  doi = {10.1103/PhysRevD.68.054022},
  url = {https://link.aps.org/doi/10.1103/PhysRevD.68.054022}
}

@article{multiple-deltas-2017,
  title = {One-dimensional semirelativistic Hamiltonian with multiple Dirac delta potentials},
  author = {Erman, Fatih and Gadella, Manuel and Uncu, Haydar},
  journal = {Phys. Rev. D},
  volume = {95},
  issue = {4},
  pages = {045004},
  numpages = {30},
  year = {2017},
  month = {Feb},
  publisher = {American Physical Society},
  doi = {10.1103/PhysRevD.95.045004},
  url = {https://link.aps.org/doi/10.1103/PhysRevD.95.045004}
}

@article{eckstein,
  title = {Causal evolution of wave packets},
  author = {Eckstein, Micha\l{} and Miller, Tomasz},
  journal = {Phys. Rev. A},
  volume = {95},
  issue = {3},
  pages = {032106},
  numpages = {13},
  year = {2017},
  month = {Mar},
  publisher = {American Physical Society},
  doi = {10.1103/PhysRevA.95.032106},
  url = {https://link.aps.org/doi/10.1103/PhysRevA.95.032106}
}

@article{pavsic,
	title = {Localized {States} in {Quantum} {Field} {Theory}},
	volume = {28},
	issn = {1661-4909},
	url = {https://doi.org/10.1007/s00006-018-0904-5},
	doi = {10.1007/s00006-018-0904-5},
	number = {5},
	journal = {Advances in Applied Clifford Algebras},
	author = {Pavšič, Matej},
	month = sep,
	year = {2018},
	pages = {89},
}

@article{meson-app-2006,
	title = {Hybrid mesons with auxiliary fields},
	volume = {29},
	issn = {1434-601X},
	url = {https://doi.org/10.1140/epja/i2006-10090-0},
	doi = {10.1140/epja/i2006-10090-0},
	number = {3},
	journal = {The European Physical Journal A - Hadrons and Nuclei},
	author = {Buisseret, F. and Mathieu, V.},
	month = sep,
	year = {2006},
	pages = {343--351},
}

@article{annalen,
author = {Torre, Amalia and Lattanzi, Ambra and Levi, Decio},
title = {Time-Dependent Free-Particle Salpeter Equation: Numerical and Asymptotic Analysis in the Light of the Fundamental Solution},
journal = {Annalen der Physik},
volume = {529},
number = {9},
pages = {1600231},
doi = {https://doi.org/10.1002/andp.201600231},
url = {https://onlinelibrary.wiley.com/doi/abs/10.1002/andp.201600231},
eprint = {https://onlinelibrary.wiley.com/doi/pdf/10.1002/andp.201600231},
year = {2017}
}

@Article{us-dynamics,
AUTHOR = {Gutierrez de la Cal, Xabier and Matzkin, Alex},
TITLE = {Beyond the Light-Cone Propagation of Relativistic Wavefunctions: Numerical Results},
JOURNAL = {Dynamics},
VOLUME = {3},
YEAR = {2023},
NUMBER = {1},
PAGES = {60--70},
URL = {https://www.mdpi.com/2673-8716/3/1/5},
ISSN = {2673-8716},
DOI = {10.3390/dynamics3010005}
}

@article{Buchleitner,
title = {Non-dispersive wave packets in periodically driven quantum systems},
journal = {Physics Reports},
volume = {368},
number = {5},
pages = {409-547},
year = {2002},
issn = {0370-1573},
doi = {https://doi.org/10.1016/S0370-1573(02)00270-3},
url = {https://www.sciencedirect.com/science/article/pii/S0370157302002703},
author = {Andreas Buchleitner and Dominique Delande and Jakub Zakrzewski},
}

@book{gradstejn_table_2009,
	address = {Amsterdam},
	edition = {7. ed., [3. Nachdr.]},
	title = {Table of integrals, series and products},
	isbn = {978-0-12-373637-6},
	language = {en},
	publisher = {Elsevier Acad. Press},
	author = {Gradštejn, Izrail S. and Ryžik, Josif M. and Jeffrey, Alan and Zwillinger, Daniel and Gradštejn, Izrail S.},
	year = {2009},
}

@book{Barut2010,
  author    = {A. O. Barut},
  title     = {Electrodynamics and Classical Theory of Fields and Particles},
  publisher = {Dover Publications},
  address   = {New York},
  year      = {2010},
  note      = {Originally published by Macmillan, 1964}
}

\end{document}